\pdfoutput=1
\documentclass[%
 reprint,
 amsmath,amssymb,
 aps,
]{revtex4-2}

\usepackage{graphicx}
\usepackage{dcolumn}
\usepackage{bm}

\usepackage[utf8]{inputenc}
\usepackage[T1]{fontenc}
\usepackage{mathptmx}
\usepackage[T1]{fontenc} 
\usepackage{amsmath,color}
\usepackage{amssymb}
\usepackage{amsfonts}
\usepackage{amsthm}
\usepackage{mathtools}
\usepackage[title]{appendix}
\usepackage{hyperref}
\usepackage{braket}
\usepackage{xcolor}
\usepackage{etoolbox}
\usepackage{soul}
\newcommand*{\citen}[1]{%
  \begingroup
    \romannumeral-`\x 
    \setcitestyle{numbers}%
    \cite{#1}%
  \endgroup
}

\usepackage{algorithm}
\usepackage{algpseudocode}
\usepackage{dsfont}

\usepackage{bbold}

\begin{document}
\preprint{AIP/123-QED}

\title{Nuclear Angular Momentum Generation in Thermally Driven Chiral Systems}
\author{Jichen Feng}
\email{jcfeng@sas.upenn.edu}
\affiliation{ 
Department of Chemistry, University of Pennsylvania, Philadelphia, Pennsylvania 19104, USA}
\affiliation{Department of Chemistry, Princeton University, Princeton, New Jersey 
 08544, USA}
\author{Ethan Abraham}
\email{ethana@mit.edu}
\affiliation{ 
Department of Chemistry, Massachusetts Institute of Technology, Cambridge, Massachusetts 02139, USA}
\author{Joseph E. Subotnik}
\email{subotnik@princeton.edu}
\affiliation{\mbox{
Department of Chemistry, University of Pennsylvania, Philadelphia, Pennsylvania 19104, USA}}
\affiliation{Department of Chemistry, Princeton University, Princeton, New Jersey 08544, USA}
\email{subotnik@princeton.edu}
\author{Abraham Nitzan}
\email{anitzan@sas.upenn.edu}
\affiliation{ 
Department of Chemistry, University of Pennsylvania, Philadelphia, Pennsylvania 19104, USA}
\affiliation{ 
School of Chemistry, Tel Aviv University, Tel Aviv 69978, Israel
}

\date{\today}

\begin{abstract}
The appearance of angular momentum in the nuclear motion of molecular systems lacking inversion symmetry under imposed thermal gradients presents a novel mechanism with potential implications for spintronics, magnetic response, and energy transport in such systems. Here we explore this phenomenon, using theoretical analysis and numerical simulations to study angular momentum generation in several driven chiral molecular models. We demonstrate that significant vibrational angular momentum can be induced under both mechanical and thermal driving, with magnitude comparable to that induced in optically driven chiral phonons. We find that generation of angular momentum is a common and general phenomenon in driven chiral structures, highlighting the role of symmetry-breaking in the (non-equilibrium) internal atomic motion of such systems. 
\end{abstract}

\maketitle
\thispagestyle{empty}

\section{Introduction}
The consequences of linear angular momentum correlation (locking) in the behavior of electronic transport in a chiral environment have been under study for some time\cite{yan2023locking,yan2023locking2,yan2024lockingreview}. More recently, the angular momentum of nuclear motions (so-called chiral phonons) has also come into focus \cite{zhu2018phononexp,ueda2023np,wang2024summary,ishito2023phononexp,zhang2015theory,zhang2019theory,CISS_interface} and the optical excitation of chiral phonons in some ionic crystals was shown to generate giant effective magnetic fields, \cite{science2023magnetic,gao2024exp,juraschek2022theory} suggesting new opportunities for spintronics and ultrafast magnetism. Observation of such motions and their manifestations where reported in several recent papers \cite{science2023magnetic,zhu2018phononexp,gao2024exp,ueda2023np} and our own work has found that such angular-linear momentum locking occurs not only in solids but in isolated chiral molecules \cite{ethan2023quantifying,jichen2024quantifying}. Interest in chiral phonons also stems from recent suggestions that such motions are implicated in the chiral induced spin selectivity phenomenon \cite{Spin-Seedback,CISS_interface,Fransson2022temperature,Fransson2020Prb}. Recent theoretical studies based on the Boltzmann transport equation suggest that such phonons could be generated in chiral solids subjected to thermal gradients \cite{therm_angular,chen2022diodetheory,off-diagonal}.

The present study investigates the latter phenomenon, generation of nuclear angular momentum in several models of chiral molecular structures subject to thermal bias. We use both analytical calculations and molecular dynamics simulations to uncover the characteristics of such angular momentum generation by thermal gradients and mechanical driving. Three chiral models are examined in order to characterize this effect as described below. While this work focuses on chiral molecular structures, we emphasize that the necessary condition for thermal angular momentum generation is the absence of inversion symmetry, which can also arise in non-chiral     structures as well \cite{therm_angular}. We find that the nuclear angular momentum induced in driven chiral molecules strongly depends on the driving characteristics, and realization under thermal non-equilibrium reflects a spectral average. 
Furthermore, we find that the nuclear angular momentum realized in a chiral molecule under thermal non-equilibrium can reach magnitudes comparable to those induced by circularly polarized light, while exhibiting robust scaling behavior and persistence in the steady state—highlighting its potential for experimental realization and application.

The remainder of this paper is organized as follows: In Sec. II we present the Hamiltonian of our molecular Models A-C. In Sec. III we present the Langevin dynamics used to simulate the heat transport in our systems, as well as an analytical formalism to describe the angular momentum response to the driving force in the limit when the system's force field is harmonic. In Sec. IV, we discuss our results pertaining to the generation of nuclear angular momentum when the chiral models examined are subject to thermal non-equilibrium as well as mechanical driving. In Sec. V, we discuss the significance of our results and compare the anticipated experimental signatures of the angular momentum generation addressed in this work to well-known optical methods of generating such angular momentum. Finally, in Sec. VI, we conclude.

\section{\label{sec:level1}Molecular Models}

The calculations described in this work were done using three models that represent different types of molecular chirality. Model A is a harmonic helical chain, earlier used by Chen \textit{et al.}\cite{chen2022diodetheory} to study the way by which the helical structure is manifested in the correlation between linear and angular momentum. Model B is a 4-atom entity whose chirality is determined by the the way bond and dihedral angles are set. Model C is the twisted double-chain structure similar to that used by us in earlier studies\cite{ethan2023chain,ethan2023quantifying,ethan2024benchmark} in which the forcefield parameters are taken to roughly mimic a polyethylene double-chain. Models A–C are described in detail below.

\subsection{\label{sec:level2} Harmonic helical chain (Model A)}
This model, introduced in Ref. \citen{chen2022diodetheory} to demonstrate linear/angular momentum locking in  the phonon spectrum of chiral systems, is a linearly connected atomic chain whose helical structure is imposed by the tensorial character of the force-constants matrix. It is defined by the Hamiltonian \cite{chen2022diodetheory}
    \begin{equation}
        \begin{aligned}
            H_A&=\sum^N_{j=1}\frac{1}{2}m\Dot{\Vec{x}}^2_j+\sum^{N-1}_{j=1}\frac{1}{2}(\Vec{x}_j-\Vec{x}_{j+1})^T\textbf{K}_{j,j+1}(\Vec{x}_j-\Vec{x}_{j+1})\\
            &+\sum_{j=1}^N \frac{1}{2}m \omega_j^2 \Vec{x}^2_j
        \end{aligned}
    \end{equation}
where $N$ is the number of atoms, $m$ is the atomic mass (taken to be equal across all atoms), and $\omega_i$ denotes the frequency of the harmonic potential to which each atom is coupled with (only atoms that are coupled to a heat bath are coupled to a well). The chirality is defined by the anisotropy of the force constant tensor $\textbf{K}_{j}$, here set to have a 3-fold rotational-translational symmetry along the $z$-axis. Specifically,\cite{chen2022diodetheory}
    \begin{equation}
        \textbf{K}_{j,j+1}=\mathbf{R}_{z}^{-1}(2\pi j/3)\mathbf{R}_{y}^{-1}(\pi/3)\textbf{K}_{0}\mathbf{R}_{y}(\pi/3)\mathbf{R}_{z}(2\pi j/3),
    \end{equation}
    with
    \begin{equation}\label{equation:K0}
         \textbf{K}_{0}=\begin{pmatrix}
            k_1 & 0 & 0 \\
            0 & k_2 & 0 \\
            0 & 0 & k_3 \\
        \end{pmatrix},
    \end{equation} 
where  $\mathbf{R}_{y}$, $\mathbf{R}_{z}$ denote rotations about the $y$ and $z$-axes, respectfully. Taking the $x$-axis as the direction of a given bond, the next bond is generated by rotating the given bond around the  $y$-axis by $60^\circ$  followed by rotation around the  $z$ axis by $120^\circ$ . Following Ref. \citen{chen2022diodetheory}, the calculations reported in Section 3 use the values  $k_1=1.0$, $k_2=0.05$, $k_3=0.25$. Note that this model is purely harmonic and as such easily amenable to analytical treatment.

\subsection{\label{sec:level2}  4-atom molecule \\
(Model B [anharmonic] and Model B$_\textit{\textbf{h}}$ [harmonic])}
In order to better understand the fundamental nature of the observed phenomenon, we next sought the simplest possible system in which it could be observed. Indeed, a 1-site system cannot exhibit heat flux, while less than 4 sites cannot exhibit chiral structure. Therefore, the simplest such candidate is a four-site system. Nonetheless, it is not immediately obvious that 4 sites require 4 atoms. One might try to look at a 2-atom system and encode chirality in directional interactions with the hot and cold baths (i.e. directional Langevin forces). Our such attempts to observe the effect in a two atom model directionally coupled to Markovian heat baths failed, while angular momentum is clearly seen to be generated in a 4-atom model. Mathematically, from the perspective of a 2-atom model with directional couling to heat baths, the only difference between the models is the non-markovian character of the thermal baths seen by the 2-atom subsystem. We leave this interesting observation to a future study.

We thus consider the 4-atom model described by the Hamiltonian 
    \begin{equation}
        \begin{aligned}
            H_{B}&=\sum_{j=1}^4 \frac{1}{2}m\Dot{\Vec{x}}^2_j+V_B\left(\{\Vec{x}_j \}\right),
        \end{aligned}
    \end{equation}
    \begin{equation}
        V_B\left(\{\Vec{x}_j \}\right)=\sum^{3}_{j=1}\frac{1}{2}K\left[|(\Vec{x}_j+\Vec{r}_j)-(\Vec{x}_{j+1}+\Vec{r}_{j+1})| -r_0\right]^2+\sum_{j=1}^4 \frac{1}{2}m \omega_0^2 \Vec{x}^2_j
    \end{equation}
where $\vec{r}_j$ are the atomic equilibrium positions, taken to satisfy $|\Vec{r}_j-\vec{r}_{j+1}|=r_0$ (equilibrium bond length). Note that the absolute positions $\Vec{r}_j$ in space define the molecular shape, while the $\vec{x}_i$ describe harmonic oscillations about these equilibrium positions with frequency $\omega_0$. In the calculations described below we use $K=500~\mathrm{N/m}$, and the frequency of the harmonic well to locate the atom is chosen as $\omega=16~\mathrm{THz}$. $m=12~\mathrm{AMU}$ and the bond length is set to be $r_0=1.5~\AA$.

The molecular shape is determined by the equilibrium vectors $\Vec{r}_j$  as follows: Start with all atoms in the $xz$-plane with the center bond, between atoms 2 and 3, along the $z$ axis and with the angle between the 1-2 and 2-3 bonds as well as the angle between 2-3 and the 3-4 bonds set to $\beta$ (so that the azimuthal angle is $\theta=\pi-\beta$). Next, keeping atom 3 fixed, rotate the 3-4 bond around $Z$-axis by angle $\alpha$ , moving atom 4 out of the $x$-$z$ plane. The angle $\alpha$ between the 1-2-3 ($x$-$z$) and 2-3-4 planes is the dihedral angle of the structure. The resulting structure, visualized in Fig. 4(a), is given by

    \begin{equation}
        \begin{aligned}
            \Vec{r}_1(\alpha,\theta)&=r_0(-\cos(\theta),0,\sin(\theta))\\
            \Vec{r}_2(\alpha,\theta)&=r_0(0,0,1)\\
            \Vec{r}_3(\alpha,\theta)&=r_0(\cos(\theta)\cos(\alpha),\cos(\theta)\sin(\alpha),\sin(\theta))\\
        \end{aligned}
    \end{equation}
Results for the dynamical behavior of this structure are shown below. However, note that the construction of this system is not limited to the one introduced, as the bond angle could be adjusted with the dihedral angle. The result of other constructions are shown in the Supplementary VI.

Note that although the interactions that characterize the potential (4)-(5) are quadratic in the deviation of each bond from its equilibrium length, the restriction imposed by the molecular shape (represented by the absolute value term in this potential) makes it anharmonic. A harmonic approximation can be derived by expanding the interaction (5) in the small deviation from equilibrium, denoted for the bond  $(j,j+1)$ by  $\delta x_{j,j+1}^\parallel$ and $\delta x_{j,j+1}^\perp$  along the bond direction and perpendicular to it, respectively. Using
    \begin{equation}
        \begin{aligned}\label{eq:expansion}
            \sqrt{(r_0+\delta x_\parallel)^2+\delta x_\perp^2} -r_0
            =\delta x_\parallel + \mathcal{O}(\delta x^2),
        \end{aligned}
    \end{equation}
this leads in the harmonic approximation to the potential
    \begin{equation}
        V_{B_h}\left(\{\Vec{x}_j \}\right)=\sum^{3}_{j=1}\frac{1}{2}K \delta {x^\parallel_{j,j+1}}^2+\sum_{j=1}^4 \frac{1}{2}m \omega_0^2 \Vec{x}^2_j.
    \end{equation}
This potential can be be recast in a form similar to Eq. (1) 
    \begin{equation}
        V_{B_h}\left(\{\Vec{x}_j \}\right)=\sum^{3}_{j=1}\frac{1}{2}(\Vec{x}_j-\Vec{x}_{j+1})^T\textbf{K}_{j,j+1}(\Vec{x}_j-\Vec{x}_{j+1})+\sum_{j=1}^4 \frac{1}{2}m \omega_0^2 \Vec{x}^2_j
    \end{equation}
 where  $\textbf{K}_{j,j+1}$ are bond force-constant tensors given by
 \begin{equation}
     \textbf{K}_{1,2}=\mathbf{R}_{y}^{-1}(\theta-\pi/2)\textbf{K}_{0}\mathbf{R}_{y}(\theta-\pi/2)
 \end{equation}
 \begin{equation}
     \textbf{K}_{2,3}=\mathbf{R}_{y}^{-1}(\pi/2)\textbf{K}_{0}\mathbf{R}_{y}(\pi/2)
 \end{equation}
 \begin{equation}
     \textbf{K}_{3,4}=\mathbf{R}_{z}^{-1}(\alpha)\mathbf{R}_{y}^{-1}(-\theta+\pi/2)\textbf{K}_{0}\mathbf{R}_{y}(-\theta+\pi/2)\mathbf{R}_{z}(\alpha)
 \end{equation}
where
    \begin{equation}
        \textbf{K}_{0}=\begin{pmatrix}
            K & 0 & 0 \\
            0 & 0 & 0 \\
            0 & 0 & 0 \\
        \end{pmatrix}.
    \end{equation}

More clearly, the potential (9) can be written in the form
\begin{equation}
    V_{B_h}({\mathbb{X}})=\frac{1}{2}\mathbb{X}^\dagger \mathbb{K} \mathbb{X}+\frac{1}{2}m\omega_0^2\mathbb{X}^2
\end{equation}
with the force constant tensor given by 

    \begin{equation}
        \mathbb{K}=\begin{pmatrix}
            -\textbf{K}_{1,2} & \textbf{K}_{1,2} & 0 & 0\\
            \textbf{K}_{1,2} & -\textbf{K}_{1,2}-\textbf{K}_{2,3} & \textbf{K}_{2,3} & 0\\
            0 & \textbf{K}_{2,3} & -\textbf{K}_{2,3}-\textbf{K}_{3,4} & \textbf{K}_{3,4} \\
            0 & 0 & \textbf{K}_{3,4} & -\textbf{K}_{3,4}
        \end{pmatrix}.
    \end{equation}

Below we refer to this harmonic approximation as Model $\mathrm{B_\textit{h}}$.

\subsection{Polyethylene double-helix (Model C)} Our third model is a double-chain polyethylene wire twisted to form a double helix as described in Ref.~\citen{ethan2023chain}. Where $N$ is the polymerization of each individual chain, the Hamiltonian is given by\begin{equation}
\begin{split}
H_C & = \sum_{j=1}^{2N}\frac{1}{2}m\dot{\vec{x}}_j^2 + \sum_{j=1}^{2N-2} k_b (l_j-l_0)^2 + \sum_{j=1}^{2N-4} k_\theta (\theta_j-\theta_{0})^2 + \\
  & \sum_{j=1}^{2N-6} \sum_{n_j}^4 \frac{C_n}{2} \left[ 1+ (-1)^{n_{j-1}} {\rm{cos}} ( n_j \phi_j) \right] + \\ & \sum_{j=1}^{2N} \sum_{i \neq j} 4 \epsilon_{ij} \left[  (\frac{\sigma}{r_{ij}})^{12} - (\frac{\sigma}{r_{ij}})^{6}  \right],
\end{split}
\label{Hmol}
\end{equation} where $\dot{\vec{x}}_j$ is the velocity of atom $j$, $l_j$,$\theta_j$, and  $\phi_j$ are the values of the $j$th bond-length, angle, and dihedral respectively, and $r_{ij}$ are the inter-atom distances. Accordingly, the parameters $k_{b}$ and $k_{\theta}$ are the spring constants for the bonds and angles, and $C_n$ are constants that define the dihedral potentials. The $\sigma$ and $\epsilon_{ij}$ are Leonard-Jones parameters to account for non-bonded interactions  (such as between the strands in opposite chains). The above parameter values were chosen as in Ref. \citen{Hadi} to fit observed physical properties \cite{Martin1998,Wick2000}.. Compared to Models A and B, this model more closely represents an experimentally observable system, though at the expense of analytical tractability. In particular, the force-constant tensor of this model does not have a concise form as in Eq.~(1) and Eq.~(9), although it could principle be computed numerically in the basis of normal mode coordinates under the harmonic approximation as was done in Ref.~\citen{ethan2023quantifying}.

\section{Theoretical and numerical calculations}

A standard numerical way to simulate a classical system coupled to one or more heat baths is Langevin dynamics. In the Markovian limit the equation of motion for atom $j$ is 
\begin{equation}
    \begin{aligned}
        m\Ddot{\Vec{x}}_j=-\frac{\partial V(\Vec{x}_j)}{\partial \Vec{x}_j}-\gamma_j m \Dot{\Vec{x}}_j+R_j(t)
    \end{aligned}
\end{equation}
with Gaussian random force $R_j(t)$ and friction $\gamma_j$ that satisfy the fluctuation-dissipation relation
\begin{equation}\label{langevin}
    \langle R_j(t) \rangle=0;~
    \langle R_j(0)R_j(t)\rangle=2m\gamma_jk_BT_j\delta(t),
\end{equation}
where $T_j$ is a constant that defines the temperature of atom $j$.

A desired thermal gradient is imposed by attaching the atoms on the opposite sides of the molecules to heat reservoirs of different temperatures. An effective temperature of atom $j$ can be defined by its ensemble averaged  kinetic energy, $K_E^j=\frac{1}{2}m\langle \Dot{\Vec{x}}_j^2 \rangle=\frac{3}{2}k_BT_j$.  The heat flux between an atom $j$ and the heat bath it is attached to can be evaluated by \cite{Sasa2006Heatconduct}

\begin{equation}\label{Eq:fluxcold}
    J_j=\gamma_j m \Dot{\Vec{x}}_j^2 -\frac{3}{2}k_BT_{bath}^{(j)}\gamma_j
\end{equation}
At steady state $\sum_j J_j=0$. In a typical implementation some atoms are connected to a cold bath and others to a hot bath, and the corresponding sums,  $J_c$ and $J_h$ satisfy $J_c+J_h=0$. The steady-state heat current through the system is then given by $J=J_c=-|J_h|$. When performing such Langevin dynamics simulations for Models A-C, we allow the system to relax for several nanoseconds until the steady-state is reached.

In addition to such Langevin dynamics simulations, the characteristics of vibrational energy flow in these systems can also be studied using a driven-damped version of the corresponding molecular models. In such calculation we apply a periodic driving force on one end of the system and damping (equivalent to a zero-temperature bath) on the other. In our implementation, for a molecular chain directed along the $z$-axis we apply an oscillating force in the $xy$-plane along a specific direction 

$F_1=(F \cos(\phi)cos(\omega t),F\cdot \sin(\phi)\cos(\omega t+\eta),0)$  on the atom on one molecular edge (atom 1, say), and damp the motion by applying friction at the other end. For better numerical stability the energy loss rate is spread over several (typically 20) atoms at the cold end of the system. In the calculation reported below we use $\eta=0$  or $\eta=\pi/2$  to represent linear (in a direction determined by $\phi$) or circular driving. The steady state of such driven system is characterized by an energy flux that depends on the amplitude, frequency, and phase of the driving force.

In executing these simulations with extended molecular chians (e.g. Models A and C) we sometimes found it useful to use a configuration in which the driven side (or the hot both) is taken at the middle of the chain with damping (or cold bath) applied at the two ends (see Fig. 1(a)). This makes it possible to evaluate the heat (energy) current going in the two opposite directions as a way to observe preferred directionality of transport along the helical chain, which will be addressed in a forthcoming article.

The Langevin dynamics simulations described above generate steady states trajectories that are used to calculate observables of interest. Our present focus is on the atomic angular momentum. For example, its $z$ component is given by

\begin{equation}\label{equation:angular_momentum_equation}
    L_{z,j}(t)=mx_j(t)\Dot{y}_j(t)-my_j(t)\Dot{x}_j(t)
\end{equation}
where $x_j$ and $y_j$ are corresponding components of the displacement of atom $j$ from its equilibrium position. An average over time
\begin{equation}\label{thermav}
    \langle L_{z,j} \rangle =\frac{1}{T}\int_0^T \left[mx_j(t)\Dot{y}_j(t)-my_j(t)\Dot{x}_j(t)\right]\mathrm{d}t.
\end{equation}
and (for the thermal simulations) over an ensemble of trajectories yields the desired observable.

The total angular momentum in the molecular system is obtained by summing over all atoms
\begin{equation}\label{Lz}
         \langle L_z \rangle =\sum_{j=1}^N  \langle L_{z,j} \rangle.
\end{equation}

The procedure described above can be applied for any molecular force field. However, when the force field is harmonic (Models A and $B_h$), the dynamical equations of motion of the driven-damped problem can be solved analytically. For simplicity we take all atomic masses equal. The equations of motion

    \begin{equation}
        \begin{aligned}
              &m\Ddot{\Vec{x}}_1=-\textbf{K}_{1,2}(\Vec{x}_1-\Vec{x}_{2})+\Vec{F}_1-m\gamma_1 \Dot{\Vec{x}}_1-m\omega_1^2\Vec{x}_1,\\
              &m\Ddot{\Vec{x}}_j=-\textbf{K}_{j-1,j}(\Vec{x}_j-\Vec{x}_{j-1})-\textbf{K}_{j,j+1}(\Vec{x}_j-\Vec{x}_{j+1})+\Vec{F}_j\\&~~~~~~~~~~~~~~~~~~~~~~~~~~~~~~~~~~~~~~~~~~-m\gamma_j \Dot{\Vec{x}}_j-m\omega_j^2\Vec{x}_j,\\
              &\quad \quad \quad \quad \vdots\\
              &m\Ddot{\Vec{x}}_N=-\textbf{K}_{N-1,N}(\Vec{x}_N-\Vec{x}_{N-1})+\Vec{F}_N-m\gamma_N \Dot{\Vec{x}}_N-m\omega^2\Vec{x}_N,          
          \end{aligned}
    \end{equation}

can be cast in the compact form 
    \begin{equation}\label{eq:compact}
        m\Ddot{\mathbb{X}}=\mathbb{K}\cdot \mathbb{X}-m\mathbb{\Gamma}\cdot \Dot{\mathbb{X}}+\mathbb{F},
    \end{equation}
    where

    \begin{equation}
        \mathbb{X}=\left (\Vec{x}_1,\Vec{x}_2, \Vec{x}_3,...,\Vec{x}_N \right )^T
    \end{equation}
    \begin{equation}
    {\scriptsize
    \mathbb{K}=
        \begin{pmatrix}
            -\textbf{K}_{1,2} &\textbf{K}_{1,2}  & 0 & 0 &...& 0 \\
            \textbf{K}_{1,2} & -\textbf{K}_{1,2}-\textbf{K}_{2,3} & \textbf{K}_{2,3} & 0 &...&0 \\
            0 & \textbf{K}_{2,3} & -\textbf{K}_{2,3}-\textbf{K}_{3,4} & \textbf{K}_{3,4} &...&0 \\
            0& 0& ...& & & \\
            0& 0& &... &\textbf{K}_{N-1,N} &  -\textbf{K}_{N-1,N} \\
        \end{pmatrix},
        }
    \end{equation}
    
    \begin{equation}
        \mathbb{\Gamma}=\mathrm{Diag}\left [\gamma_1,\gamma_1,\gamma_1,...,\gamma_j,\gamma_j,\gamma_j,...,\gamma_N \right],
    \end{equation}
and
    \begin{equation}
        \mathbb{F}=\left[ \Vec{F}_1, \Vec{F}_2, \Vec{F}_3,...,\Vec{F}_N \right]^T.
    \end{equation}

The steady state reached while driving with frequency $\omega$ is described by the Fourier transform of Eq.~(\ref{eq:compact})
    \begin{equation}        
    -m\omega^2\Tilde{\mathbb{X}}=\mathbb{K}\cdot\Tilde{\mathbb{X}}-im\omega\mathbb{\Gamma}\cdot \Tilde{\mathbb{X}}+\Tilde{\mathbb{F}},
    \end{equation}

Here the tildes denote Fourier conjugates in frequency space, e.g $\tilde{\mathbb{X}}=\tilde{\mathbb{X}}(\omega).$ The solution to Eq.~\ref{eq:compact} is
\begin{equation}
\label{equation:FT_X}
    \Tilde{\mathbb{X}}(\omega)=\tilde{\mathbb{G}}(\omega)\Tilde{\mathbb{F}}(\omega)
\end{equation}
where $\tilde{\mathbb{G}}$ is the Green’s function
    \begin{equation}\label{greens}
        \tilde{\mathbb{G}}(\omega)=[-\mathbb{K}+im\omega\mathbb{\Gamma}-m\omega^2\mathbb{1}]^{-1}.
    \end{equation}

For an arbitrary driving force $\mathbb{F}(t)$, its Fourier transform $\tilde{\mathbb{F}}(\omega)$ has Cartesian components
\begin{equation}\label{Eq:components}
F_{j\sigma}(\omega) = \Tilde{F}_0(\omega)\tilde{f}_{j\sigma}(\omega),
\end{equation}\label{Eq:monochrom} where $\Tilde{F}_0(\omega)$ stands for the amplitude of the force and is real while $\tilde{f}_{j\sigma}(\omega)$, satisfying $\sum_\sigma|\tilde{f}_{j\sigma}(\omega)|^2=1,$ gives the vector specifying the direction and phase of the force on each atom for \(j = 1,\ldots,N \) and \( \sigma \in \{x, y, z\} \) (in our model, $j$ only equals to 1, indicating our system is connected to the hot bath with one edge atom).

The motion of atom $j$ due to one such monochromatic component is given by
    \begin{equation}
    \label{equation:XY}
        \begin{aligned}
            x_j(t)&=(A_je^{i \omega t}+A_j^*e^{-i \omega t})\Tilde{F}_0(\omega)\\
            y_j(t)&=(B_je^{i \omega t}+B_j^*e^{-i \omega t})\Tilde{F}_0(\omega)\\
            z_j(t)&=(C_je^{i \omega t}+C_j^*e^{-i \omega t})\Tilde{F}_0(\omega)\\
        \end{aligned}
    \end{equation}
where $\Tilde{F}_0(\omega)$ is the force amplitude and $A_j$, $B_j$ and $C_j$ are calculated from Eq.~(\ref{equation:FT_X}).

The angular momentum along $z$-axis of atom $j$ at time $t$ is readily obtained from Eq.~\ref{equation:angular_momentum_equation} and Eq.~(\ref{equation:XY}). The average steady state value is obtained from Eq. (\ref{thermav}) with $T=2\pi/\omega$
    \begin{equation}\label{Eq:Lzj}
        \begin{aligned}
            L_{z,j}(\omega)=4m\omega \mathrm{Im}[A_jB_j^*]\tilde{F}_0(\omega)^2.
        \end{aligned}
    \end{equation}

Similarly for the heat current, plugging in $T_{cold}=0$ in Eq.~\ref{Eq:fluxcold}, using Eq.~\ref{equation:XY}, and averaging over time and phase with frequency $\omega$ leads to 
\begin{equation}\label{Eq:J}
    \begin{aligned}
         J_j=2\omega^2 m\gamma_j(|A_j|^2+|B_j|^2+|C_j|^2)\tilde{F}_0(\omega)^2.
    \end{aligned}
\end{equation}

Alternatively, we can define the angular momentum operator $\hat{L}_z^j$ of atom j and the total angular momentum $\hat{L}_z$ for the system by 
\begin{equation}
    \hat{L}_z^j=\begin{pmatrix}
                    0 & -i & 0 \\
                    i & 0 & 0 \\
                    0 & 0 & 0 \\
                \end{pmatrix}^{(j)};\quad \hat{L}_z=\bigoplus_{j=1}^N \hat{L}_z^j
\end{equation}
The latter then takes the form 
\begin{equation}\label{Eq:Lz_fourier}
    L_z(\omega)=-\frac{1}{2}m\omega\tilde{\mathbb{X}}^\dagger\hat{L}_z\tilde{\mathbb{X}}=-\frac{1}{2}\sum_{j=1}^Nm\omega \tilde{\mathbb{X}}^\dagger_j \hat{L}_z^j \tilde{\mathbb{X}}_j
\end{equation}
where $\tilde{\mathbb{X}}_j$ is a Cartesian vector with components $\tilde{X_i},\tilde{Y_i},\tilde{Z_i}$. 

The total angular momentum can also be expressed in a compact kernel form that highlights its dependence on the driving field. Inserting Eq.~(\ref{equation:FT_X}) into Eq.~(\ref{Eq:Lz_fourier}) yields
\begin{equation}
\label{Eq:response_kernel}
L_z(\omega) = -\frac{1}{2} m \omega \, \tilde{\mathbb{F}}^\dagger \, \tilde{\mathbb{G}}^\dagger \hat{L}_z \tilde{\mathbb{G}} \, \tilde{\mathbb{F}},
\end{equation}
where $\tilde{\mathbb{G}}(\omega)$ is the system Green’s function defined above. Eq.~\ref{Eq:response_kernel} explicitly shows that $L_z(\omega)$ is a quadratic form in the driving field, and the response kernel $\tilde{\mathbb{G}}^\dagger \hat{L}_z \tilde{\mathbb{G}}$ encodes how system symmetry and coupling mediate rotational motion under external driving.

To connect this analysis with simulation, we consider the case of driving with frequency $\omega_0$ on one atom of one end of the chain, while the heat bath on the other end of the chain is set to $T_\text{cold}=0,$ corresponding to damping without driving.(Note that for the simulation result, we choose difference temperature for the cold bath, i.e. $T_{cold}=2~\mathrm{K}$) In the Fourier space of our system, the form of such a force is, e.g.
\begin{equation}\label{eq:Fdrive}
    \Tilde{\mathbb{F}}=\left[ F_0 \cos(\phi)\delta(\omega-\omega_0),F_0 \sin(\phi)\delta(\omega-\omega_0),0,...,  0, 0, 0 \right]^T,
\end{equation}
which is obtained when the external force is applied to the system in the $xy$-plane perpendicular to the chain direction, making an angle $\phi$ with the $x$-axis. For reported values, we average out the contribution of this angle (see Supplemental Sec. IV-V). The steady state motion under this driving is the inverse Fourier transform of the Eq.~(\ref{equation:FT_X})

    \begin{equation}
        \begin{aligned}\label{eq:steady}
            \mathbb{X}(t)_{Steady}
            &=\text{Re} \left\{ e^{i \omega_0 t}\mathbb{G}(\omega_0) \left[ F_0\cos(\phi),F_0 \sin(\phi),0,...,  0, 0, 0 \right]^T \right\}.
        \end{aligned}
    \end{equation}

The results of this calculation reported in Sect. IV are averaged over the angle $\phi$.The analytical solution of a general driven-damped coupled oscillators model may be extended to the case of frequency resolved thermal driving model by introducing stochastic driving forces characterized by a flat spectral density and randomized phase coherence. In this case, we obtain from Eq. (\ref{langevin})
\begin{equation}
    \langle F_0(t)^2 \rangle=\frac{m\gamma k_B T}{\pi}=\frac{1}{2}\Tilde{F}_0(\omega)^2, 
\end{equation} which can be used in Eq. (\ref{Eq:Lzj}) and Eq. (\ref{Eq:J}).

As shown in the Supplemental Sec. V, the expectation of the angular momentum averaged over all frequencies and driving angles in three dimension can be expressed as
    \begin{equation}\label{Eq:Analytical}
        \begin{aligned}
            \langle L_z\rangle&=\frac{1}{3}\int_0^\infty[L_z(\alpha=0,\theta=0,\omega_0)+L_z(\alpha=\pi/2,\theta=0,\omega_0)\\
            &+L_z(\theta=\pi/2,\omega_0)]\mathrm{d}\omega_0
        \end{aligned}
    \end{equation}
In Eq.~\ref{Eq:Analytical} $\alpha$ and $\theta$ represent the spherical coordinate of the force orientation. The total angular momentum is effectively the sum of the contribution of the driven force along $x$, $y$ and $z$ direction.(Other simplified conditions are discussed in SI~II and SI~III)

In the following section we present our results, applying the dynamics described in this section to the Models A-C described in Sec. II. In all numerical simulations performed, the dynamics are time-evolved using standard molecular dynamics.

\section{\label{sec:level3}Results}
\subsection{Harmonic helical chain}
\subsubsection{White Noise Bath}
\begin{figure*}
\includegraphics[width=480 pt]{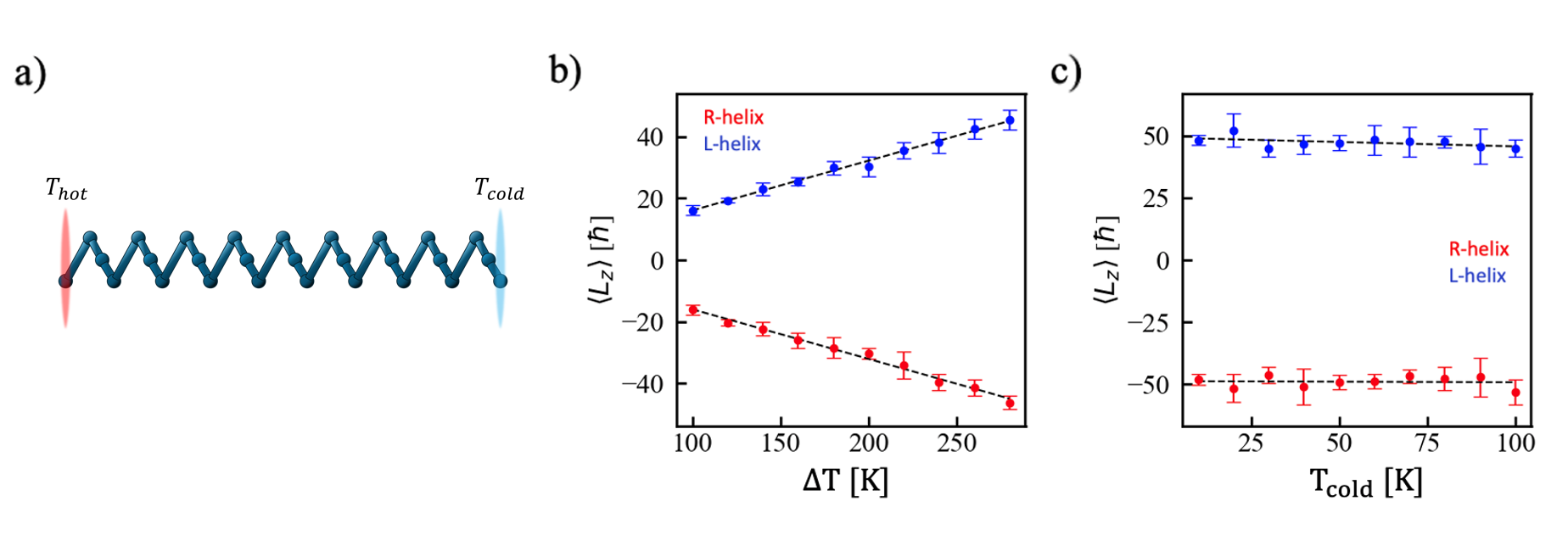}
\label{fig:1} 
\caption{
(a) Schematic of the harmonic helical chain (Model A) connected to heat baths. Setting the contact with the high-temperature bath at the one end of the chain and the contact with low-temperature baths at the other end (in the supplementary material section VII we connect the middle of the chain to the high-temperature bath and the both ends of the chain to the low-temperature bath).~
(b-c) Simulation results for Model A with polymerization $N=201$: In (b), the average angular momentum (summed over all chain
atoms) displayed against the temperature difference $\Delta T=T_{hot}-T_{cold}$ between the high and low-temperature reservoirs, where $T_{cold}$ is kept at $T_{cold}=2~\mathrm{K}$. In  (c), the total angular momentum displayed against $T_{cold}$ with $\Delta T$ kept constant at $\Delta T=300~\mathrm{K}$. Results are shown for (blue) left-handed
and (red) right-handed helical enantiomers.}
\end{figure*}

Fig.~1(a) shows a schematic of the setup used in the simulations for Model A. The parameters used for the simulation are $m=12~\mathrm{AMU}$ and $\omega=7\times10^{12}~\mathrm{s}^{-1},$ corresponding to a bond force-constant  $K=1~\mathrm{N/m}$. As described in Sec. III, a hot bath was connected to one atom at the chain edge (or center), and the cold bath was set to connect to the last few atoms (typically 10) at the cold edge(s) \footnote{This setup effects efficient damping of energy at the cold edge and is particularly important as a tool to avoid unphysical reflections when pure damping ($T=0$) is used to mimic energy flow in an infinite chain.}. The friction coefficient was taken $\gamma=3.5\times 10^{11}~\mathrm{s}^{-1}$ at all temperatures and the numerical integration was done using Verlet algorithm method with a timestep $\Delta t=14~\mathrm{fs}$ \footnote{While $\Delta t=14~\mathrm{fs}$ is higher than that of typical molecular dynamics simulations, it is justified here because in Model A it satisfies $\Delta t\simeq0.1/\omega_{max}$ where $\omega_{max}$ is the frequency of the interatomic potential.}. The local temperature of atom $j$ is evaluated from $T=\frac{m\langle v^2_j\rangle}{3K_B}$, and the angular momentum is calculated from Eq.~(\ref{Lz}).

The results for the generated angular momentum are shown in Fig. 1(b) as a function of the temperature difference between the cold and hot edges, and in Fig. 1(c) as a function of the lower temperature under constant temperature difference $\Delta T=T_{hot}-T_{cold}$. The results exhibit a clear enantiomer-dependent mean angular momentum response. It is seen that the generated angular momentum strongly depends on the temperature difference $\Delta T$ that induces the heat current $J_H$ (for harmonic chains $J_H$ depends linearly on  $\Delta T$ \cite{therm_angular}) and only weakly on the average temperature. Each data point represents an average of five independent MD trajectories and the error bars in panels (b,c) represent standard deviations from this average. 

This observation appears to be consistent with a recent theoretical consideration of this phenomenon. Using analysis based on the Boltzmann transport equation, Hamada \textit{et al.} have concluded that the steady state angular momentum generated in chiral systems is proportional to the thermal gradient \cite{therm_angular}. This conclusion is also compatible with the standard linear response theory of heat transport. We note however that in systems where phonons carrying angular momenta can be excited, arguments based on phonon angular/linear momentum locking \cite{yan2023locking,yan2023locking2,yan2024lockingreview} indicate that the primary source of net angular momentum is not the temperature gradient but the heat current. This distinction becomes important when standard linear response does not hold locally due to the delocalized nature of phonons.

An example of this distinction is shown in Fig.~2. Here the helical chain examined in Fig. 1 connects between $T_{hot}=202~\mathrm{K}$ (black) or $T_{hot}=102~\mathrm{K}$ (red) and $T_{cold}=2~\mathrm{K}$ Langevin baths. The calculated steady state heat currents are $J=1.15\times10^{-9}~\mathrm{J/s}$ and $J=5.75\times10^{-10}~\mathrm{J/s}$ for the black and red curves respectively. Fig.~2(a) shows the steady state local temperature profiles along the chain, while Fig.~2(b) displays the corresponding position dependence of the local angular momentum. It is seen that except near the edges (where the dynamics is affected by the boundary motional restrictions), the steady sate angular momentum increases with the heat flux while the thermal gradient remains nearly zero. Further study of the directionality of the heat flux coupling with the polarization of the angular momentum will be discussed in our next work.

The thermal baths used to generate the results in Fig. 1 and 2 are represented by white-Langevin noise and constant damping that are used to model classical Markovian thermal environments. Alternatively we can analyze the spectral resolution of the effect by considering the response to driving at certain frequencies. Such driving at frequency $\omega$ can be implemented by weakly connecting the system to a Markovian thermal bath via a harmonic oscillator of frequency $\omega$.

The results of such a calculation are shown in Fig. 3. First consider the fully resolved spectrum of the angular momentum generated in the molecular chain of Model A. Fig. 3(a) displays the total steady state angular momentum obtained by driving an edge atom on one side with a harmonic force perpendicular to the chain axis and imposing damping on a selected number 10 of atoms at the opposite edge. It was produced using the analytical solution, Eqs.~(\ref{equation:FT_X})-(\ref{Eq:Lz_fourier}), followed by averaging the angular momentum Eq. (\ref{Eq:Analytical}) over the driving angle and summing over frequencies. We see that not only the amplitude but also the sign of the driving-induced angular momentum generation strongly depends on the driving frequency. This dependence reflects the frequency dependent density of phonon modes of our helical chain as well as the sign of their group velocity $v_g=\frac{d\omega(k)}{dk}$ (see Fig. 3(c-d) and Ref.~\footnote{The three regimes of large response in Fig. 3a roughly correspond to the three regimes ($\omega\in(0,2.8)~\mathrm{THz}$, $\omega\in(3.8,5.2)~\mathrm{THz}$, $\omega\in(8,11)~\mathrm{THz}$) of large density of modes in Fig. 3(d), while the changing sign of the angular momentum response reflects the polarization under certain frequency region and with certain sign of group velocity in Fig.~4(c)}). This implies that the thermal-gradient induced angular momentum shown in Figs. 1 and 2 reflect partial cancellation between different frequency regimes, suggesting that a much stronger signal can be obtained in the thermal driving case by using a frequency-filtered thermal bath. 

\subsubsection{Colored Noise Bath}

Next, rather than connecting the system with white noise bath, consider a calculation for the frequency-filtered thermal bath obtained by placing a harmonic oscillator of a particular frequency $\Omega$ between the system and the white bath (colored noise bath). This effectively subjects the chiral system to a thermal bath with a power spectrum determined by the coupling between the oscillator and the bath, thus allowing us to address different spectral regimes of its response to thermal gradients. Fig. 3(b) shows results of this calculation (See the derivation of effective Langevin force in Supplementary~VII). The similarity to Fig.~3(a) is evident. Importantly, the generated angular momentum is considerable at some frequency regimes. For example, at the low frequency peak, the generated angular momentum over the heat flux is a scale higher than the corresponding value observed in Fig.~1(b) and Fig.~1(c). 

\begin{figure}
\includegraphics[width=240pt]{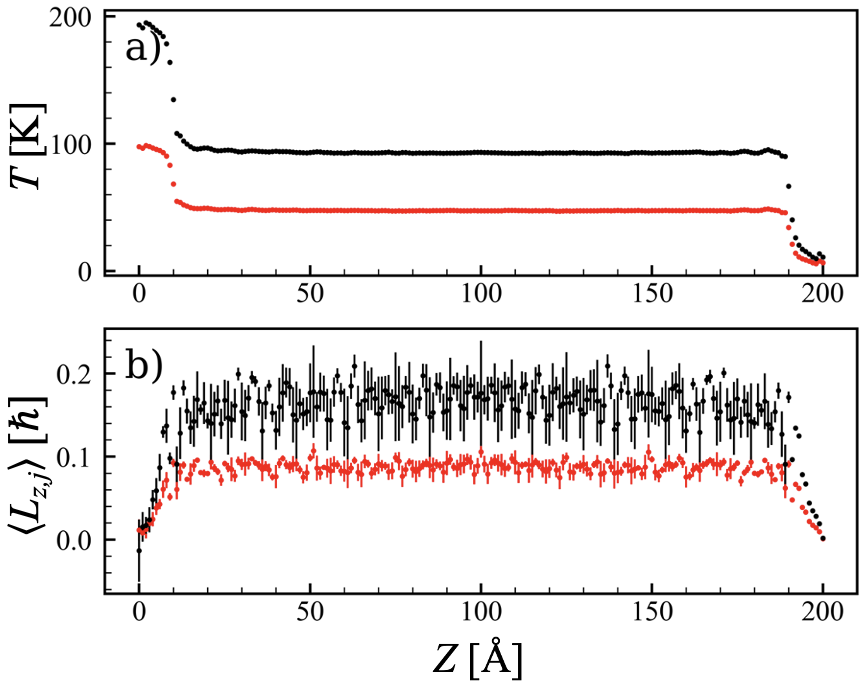}
\caption{
Langevin simulations for the molecular chain of Model A under thermal bias ($T_{cold}=2~\mathrm{K}$; see other parameter on text), showing the temperature distribution along the  $z$-axis in panel (a) and the angular momentum distribution along  $z$-axis in panel (b). The red and black lines correspond to $T_{hot}=102~\mathrm{K}$ and $T_{hot}=202~\mathrm{K}$ respectively. }
\label{fig:2}
\end{figure}

\begin{figure}
\includegraphics[width=240pt]{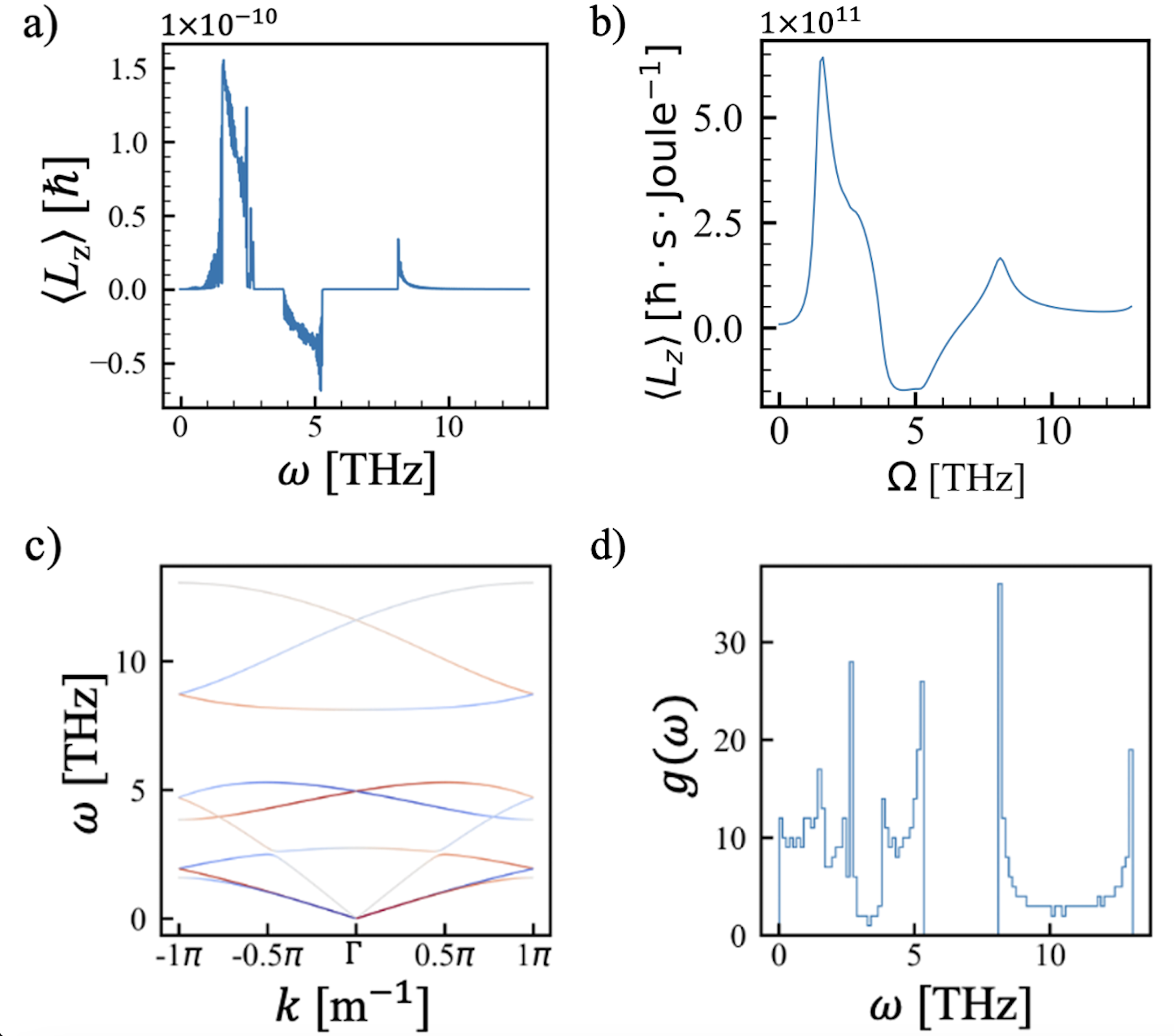}
\caption{
(a) The average total angular momentum generated by phase-averaged harmonic driving ($m=12~\mathrm{AMU}$, $T_{hot}=300~\mathrm{K}$, $T_{cold}=0~\mathrm{K}$, $\gamma=3.5\times10^{11}~\mathrm{s}^{-21}$) displayed as a function of the driving frequency. (b) Normalized total angular momentum generated in the system by using a frequency-filtered thermal bath, divided by the heat flux $(\mathrm{Joule/s})$ through the chain. The coupling constant with the bath is $\lambda=0.01~\mathrm{N/m}$. (c) Phonon band structure with the phonon angular momentum polarization\cite{zhang2014theory} labeled by color (right handed polarization labeled by red, otherwise by blue).
(d) Density of modes of the helical system. Horizontal axis is frequency and the vertical axis is the count of modes.}
\label{fig:real3}
\end{figure}

\begin{figure}
   \includegraphics[width=240pt]{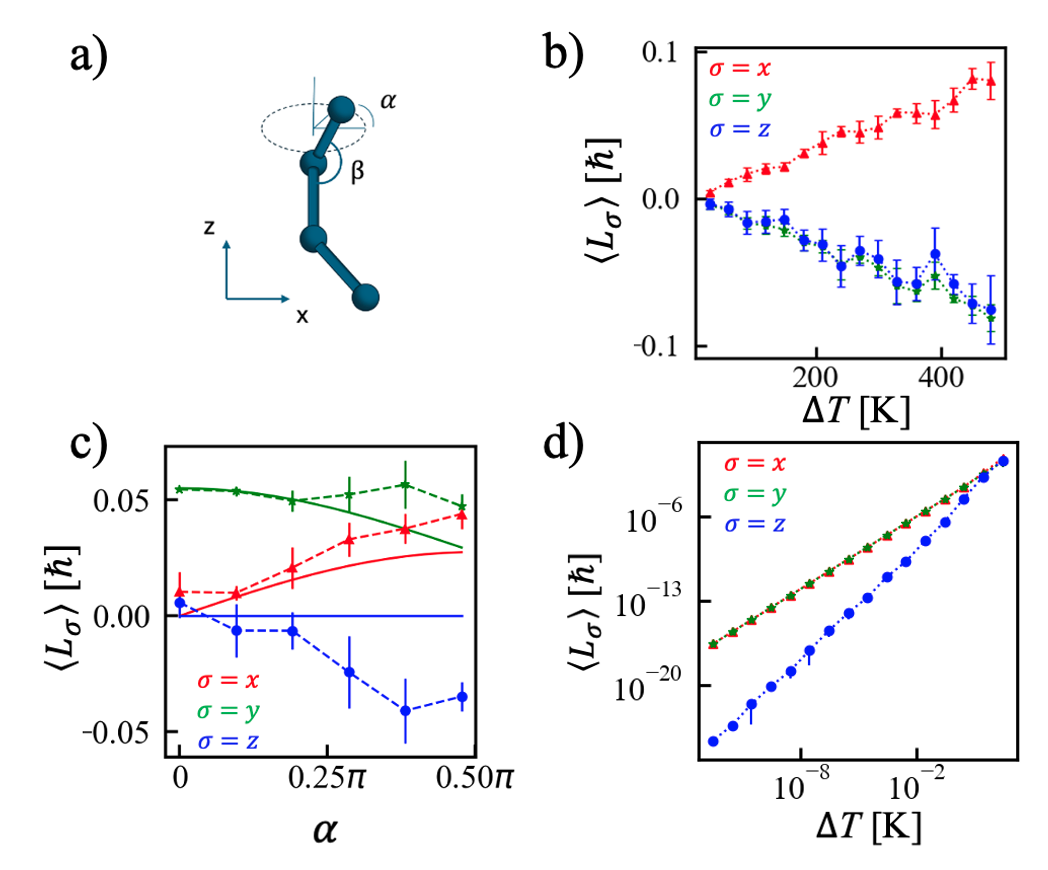}
\caption{
(a) Schematic of the 4-atom model (Model B). The equilibrium positions of three atoms are in the $x-z$ plane with an angle $\theta$ between two nearest neighbor bonds, and $\alpha$ is the dihedral angle; (b) The angular momentum component $L_\sigma$ for $\sigma =$ $x$ (red), $y$ (green), and $z$ (blue) as a function of the temperature difference $\Delta T = T_{hot}-T_{cold}$ keeping the dihedral angle fixed at $\alpha=\pi/2$. (c) The same angular momentum components plotted against the dihedral angle $\alpha$ for $T_{hot}=300~\mathrm{K}$,  $T_{cold}=0~\mathrm{K}$. Solid line: analytical calculations using the harmonic approximation, Eq.~(\ref{Eq:Analytical}). Dashed lines with error bars: numerical calculations without the harmonic approximation. (d) The same three components of the angular momentum plotted against small temperature difference keeping the dihedral angle fixed at $\alpha=\pi/2$. Shown numerical results obtained using Langevin simulations with the full potential (dashed lines with "I" error bars). Angular momentum is summed over only the two non-terminal atoms. 
}
\label{fig:4}
\end{figure}

\subsection{4-atom molecule}

More insight into the generation of nuclear angular momentum in mechanically or thermally driven molecules may be obtained by considering the 4-atom molecular model described in Section IIB. Fig.~4(a) shows a schematic of the molecular structure. The two edge atoms are coupled to heat baths of temperature $T_{cold}$ and $T_{hot}$. The random forces and damping are applied isotropically, the friction parameter was taken to be $\gamma=0.04~\mathrm{fs}^{-1}$, and the timestep used in the numerical simulations is $\Delta t=0.025~\mathrm{fs}$ .

Fig.~4(b) shows the $x$, $y$, and $z$ components of the angular momentum obtained at steady state, plotted against the temperature difference $\Delta T=T_{hot}-T_{cold}=T_{hot}$ in the $T_{cold}=0~\mathrm{K}$ limit. Fig.~4(c) shows the same components of the angular momentum with $T_{hot}=300~\mathrm{K}$ and $T_{cold}=0~\mathrm{K}$ plotted against the dihedral angle $\alpha$, with the azimuthal angle  kept fixed at $\theta=\frac{1}{4}\pi$. Fig.~4(d) shows similar results calculated for $\alpha=\pi/2$ and plotted against an infinitesimal temperature difference. When analyzing the dependence of the steady state angular momentum on $\Delta T$, we have used linear regression to fit this dependence to a power law. The following observations are noteworthy:

\begin{quote}(i) Like with the harmonic chain of Model A, in the 4-atom model we also observe that the phonon angular momentum is roughly proportional to the magnitude of the thermal gradient, as shown in Fig. 4(b). Notably, we observe the effect not only for the axial direction ($L_z$) but also for the transverse directions ($L_x,L_y$) as well.\\
(ii) Fig. 4(c) shows that chirality is not an absolute requirement for the excitation of nuclear angular momentum under simple driving: At steady-state, $L_y$ (and, in the anharmonic model, also $L_x$) remain finite even for achiral (planar) molecular structure. The observation is consistent with the observation in Ref. \cite{therm_angular} that absence of inversion symmetry, a weaker requirement than chirality, can be give rise to such angular momentum generation.\\
(iii) Fig. 4(d) shows that $L_x$ and $L_y$ are linear in $\Delta T=T_{hot}\sim0~\mathrm{K}$ for small thermal gradients. However, while we find that in the harmonic approximation (Model B$_h$) $L_z$ remains zero, we find from the numerical simulation of the full anharmonic Model B with our chosen model parameters that $L_z\sim T_{hot}^{1.55}$ in the limit $\mathrm{T_{cold}=0~K}$ and $\mathrm{T_{hot}\to0~K}$.\\
(iv) The above observations show that the magnitude of the angular momentum developed in the atomic motion of isotropically driven molecule can depend qualitatively on the molecular anharmonicity.
\end{quote}

\subsection{Polyethylene double helix}
\begin{figure}[!b]
\includegraphics[width=240pt]{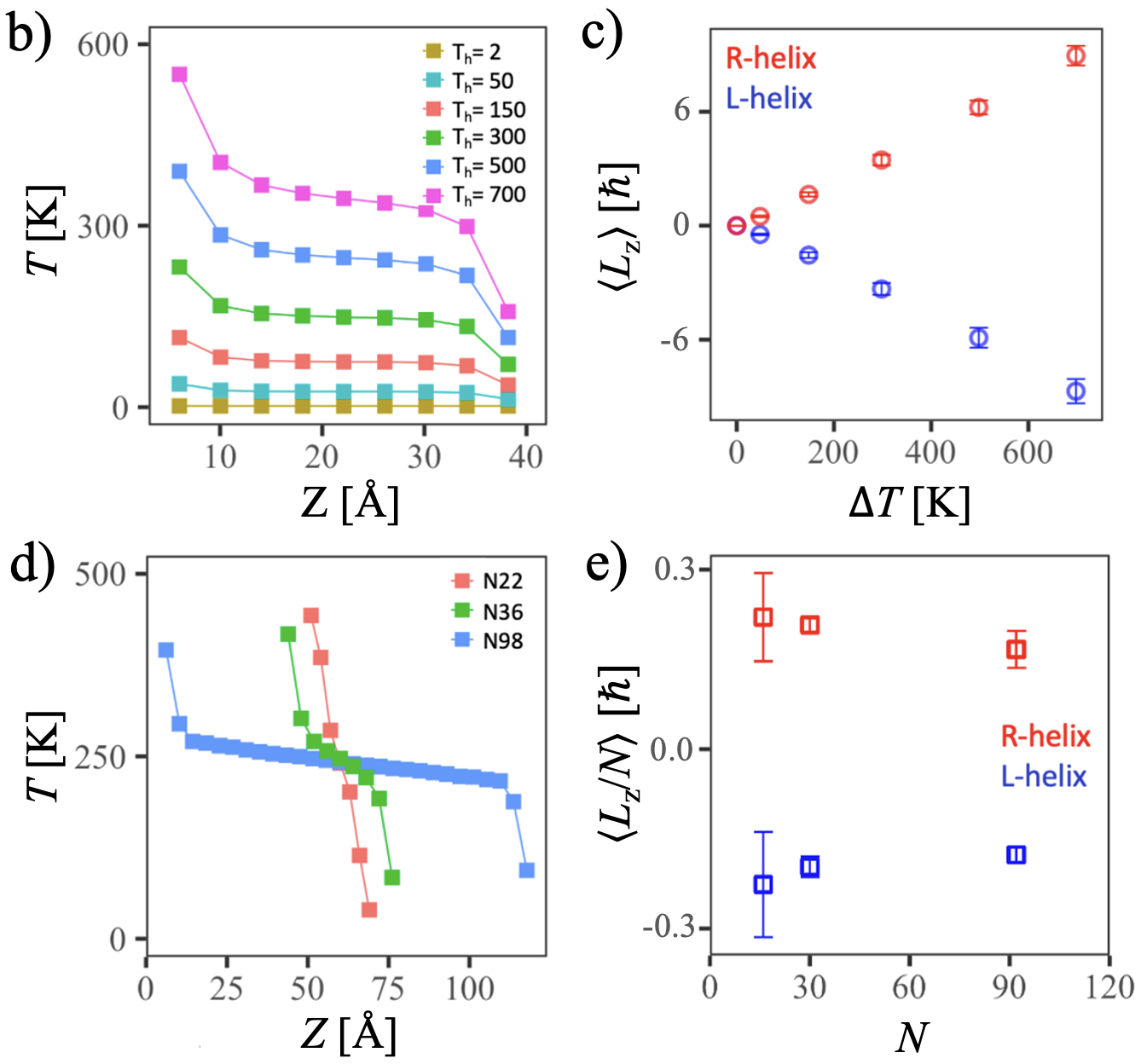}
\caption{(a) Schematic of the twisted polyethylene double-chain (Model C). (b) Average temperature profiles along the $z$-axis of an $N=36$ double-helical polyethylene wire containing 5 twists for various thermal gradients. $T_h=T_{hot}$ represents the temperature setting of the hot Langevin thermostat; in all cases the cold thermostat is set to $T_{cold}=2$ K. The colors in the figure represent $T_{hot}=2,50,150,300,500,700$~K (gold, turquoise, red, green, blue, purple respectively). (c) Average phonon angular momentum (summed over all chain
atoms) displayed against $\Delta T = T_{hot}-T_{cold}$ for a right-handed (red) and left-handed (blue) $N=36$ polyethylene double-helix with 5 twists. (d) Average temperature profiles along $N=22$ (red), $N=36$ (green), and $N=98$ (blue) double-helical polyethylene wires with the number of twists in each case chosen so that the pitch length is roughly equal to that in panels (a-c). The thermostats are set to $T_{hot}=500$, $T_{cold}=2$ K. (e)  The average phonon angular momentum per-atom shown as a function of the chain length. The error bars show the uncertainty in the averaged (over 10 ns trajectory) values obtained from 3-5 independent simulations. In (c) and (e), results are shown for a right-handed (red) and left-handed (blue) double-helical enantiomers.
}
\label{fig_PE}
\end{figure} 
A schematic of our polyethylene double helix model is shown in Fig. 5(a) and further details are described in Ref. \cite{ethan2023chain,ethan2023quantifying}. The steady-state Langevin-dynamics of this system (Section 2C) under thermal
driving are propagated with $\Delta t=1.25$ fs. Results for the generated angular momentum are shown in Fig. 5-7. 
\begin{figure}[h!]
\includegraphics[width=240pt]{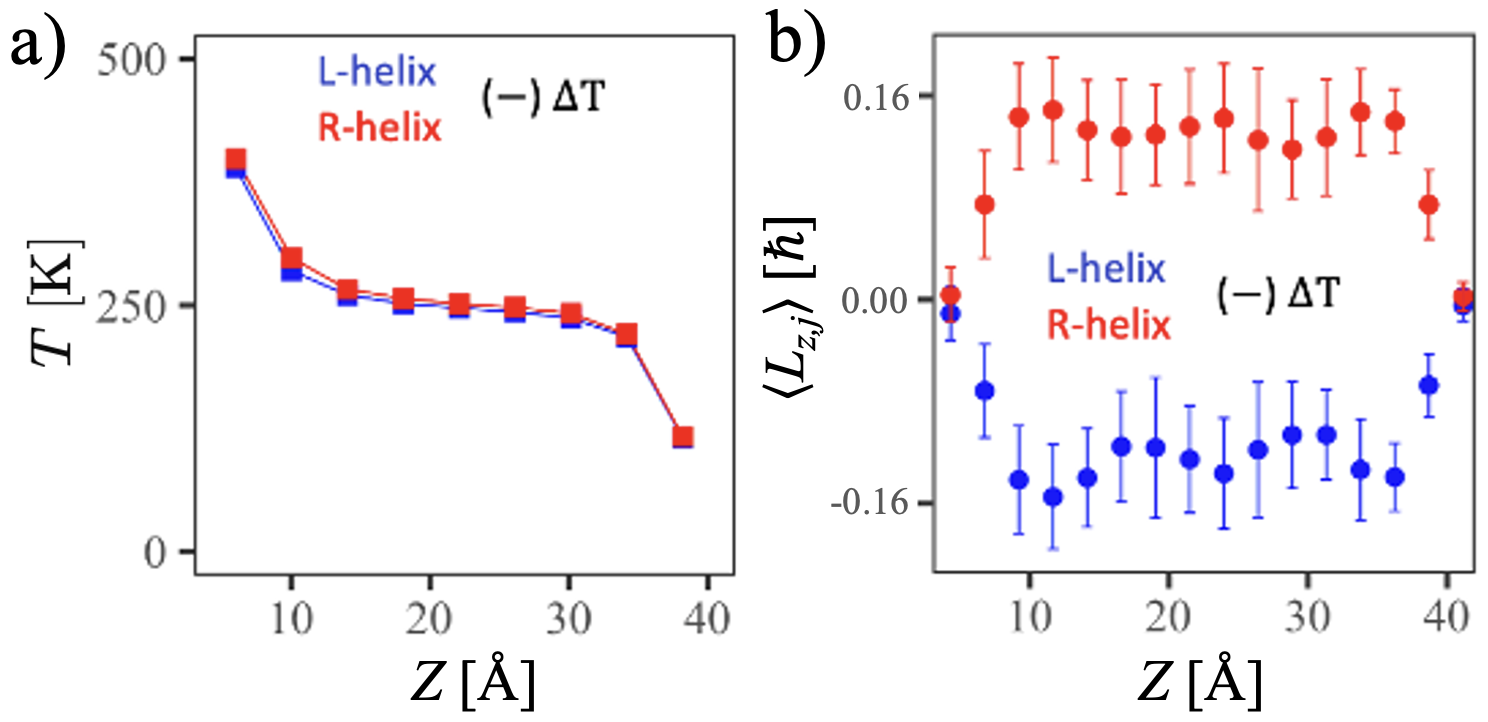}
\caption{
(a) Average temperature profiles along an $N=36$ double-helical polyethylene wire containing 5 twists for opposite handed helices. $T_{hot}$ is set to 500 K, and $T_{cold}$ to 2 K. (b) Angular momentum distribution along z-axis. Results are shown for (blue) left-handed and (red) right-handed double-helical enantiomers.
}
\label{fig_PE2}
\end{figure}The right panels (c,e) in Fig. 5 show the steady state angular momentum as function of $\Delta T = T_{hot}-T_{cold}$ ($T_{cold}$
is taken fixed at 2 K in all panels), and chain lengths $N$, while the left panels (b,d) show the corresponding temperature distribution along the chain. Figure 6 displays the temperature and angular momentum distribution along the thermally biased chain. The error bars denote the
standard deviation calculated for the (trajectory-averaged) angular momentum over three-four independent 10ns simulations. The following observations should be
noted:

\begin{quote}(i) Nuclear rotational motion is developed in a the double helix structure subjected to
thermal imbalance. For a given direction of heat flow, the sign of the angular momentum is
inverted upon inversion of the chain handedness.\\
(ii) As seen in Fig. 5(c), the total angular momentum appears to be linear in the temperature difference between the molecular edges. This linear
dependence on the temperature imbalance was also seen above in the other models studied, suggesting a general mode of behavior, at least when the transport is dominated by the harmonic part of the nuclear potential surface. However, as seen in Fig. 6, its dependence on position reflects the induced energy current, as well as local motional
restrictions near the chain edges, rather than the local temperature gradient.\\
(iii) For a given temperature difference between the edges, the steady-state angular
momentum depends only weakly on the chain length (Fig. 5e), again showing weak dependence on the temperature gradient and suggesting a more general dependence on the heat current. 
\end{quote}
\begin{figure*}[htbp]
\includegraphics[width=440pt]{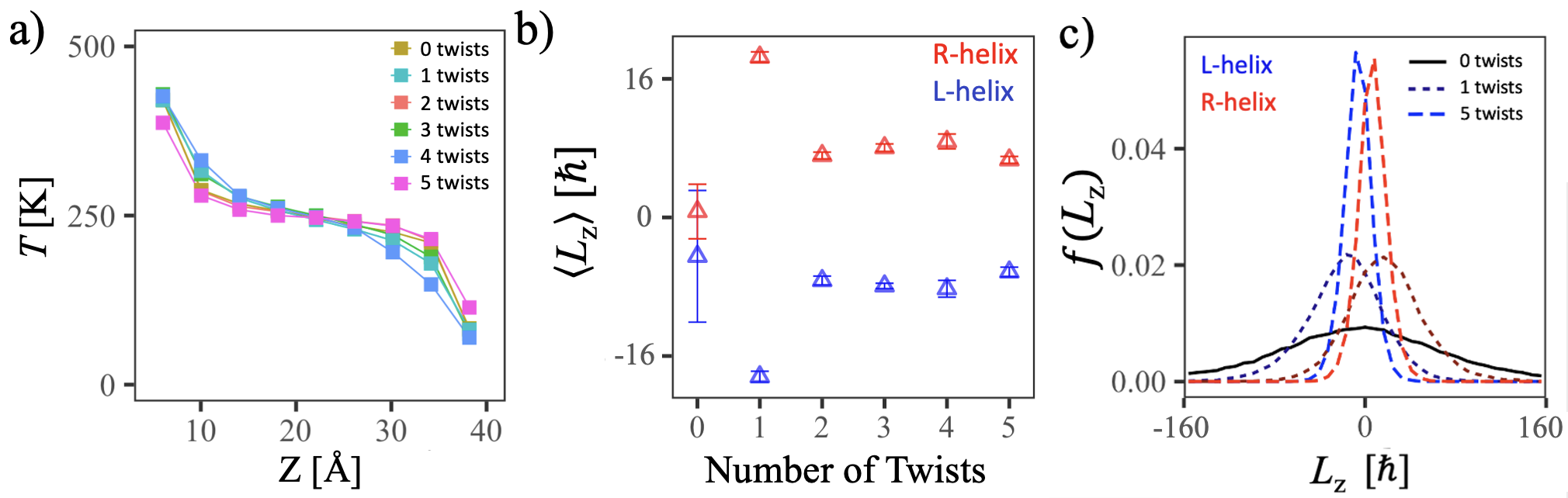}
\caption{
(a) Average temperature profiles along an $N = 36$ double-helical polyethylene wire containing 0,1,2,3,4,5,6 (gold, turquoise, red, green, blue, purple respectively) twists. The thermostats are set to $T_{hot}=500$ K, and $T_{cold}=2$ K. (b) Average phonon angular momentum (summed over all chain
atoms) as a function of the number of twists in the in the structures from (a). (c) Occurrence probability distribution of different $L_z$ values throughout the same simulations for (solid line) 0, (dotted line) 1 and (dashed line) 5 twists.}
\label{fig_PE3}
\end{figure*}

Finally, we examine the dependence of the generated angular momentum on the system's intrinsic chirality. The dependence of the average angular momentum on the number of twists in the thermally-biased double-helical chain of 2x36 beads is shown in Fig. 7(b), while the corresponding temperature profiles are shown in Fig. 7(a). We see that the angular momentum increases strongly upon the formation of a single twist, followed by a mild variation when applying further twists. This phenomenon probably reflects the opposing effects of increasing chirality on one hand, and increasing rigidity
on the other \cite{ethan2023chain}. The latter trend is evidenced by the increasing ballistic
nature of the heat transport (with a larger temperature drop near the edges) seen with increasing twists. 

Finally, we note that while this work has primarily reported the mean angular momentum values generated, further information is provided by examining the occurrence frequency of angular momentum values throughout the simulation. Figure 7(c) compares the shape of the resulting distribution for several amounts of twist. We note that the distribution of angular momentum values that occurs throughout the simulations is much broader for the untwisted structure and increases sharply with increasing twist. Furthermore, we find that the variance of the distribution is large enough such that there is still a significant probability of finding an angular momentum value with the opposite sign at any particular moment in time, suggesting that the observed angular momentum in response to biased Langevin driving is a highly incoherent response.

\section{Discussion}

Taken together, the studies described above suggest that the development of angular nuclear motion in chiral molecules under thermal imbalance is a general phenomenon with
similar characteristics in different systems. This statement is further strengthened when combined with Hamada \textit{et al.}'s study \cite{therm_angular}, which predicted a similar effect in chiral crystals such as Quartz and Tellurium. It is worth emphasizing, however, that the latter work arrived at this conclusion using the Boltzman transport equation to model a small deviations from thermal equilibrium, thus missing some important aspect of the dynamics.

In order to assess the physical significance of the nuclear angular momentum response observed above, we ask to what extent we can approximate the strength of the magnitude of the magnetic field that might arise from it. In particular, Ref. \cite{therm_angular} predicted that this effect would be small under the assumption of fixed Born charges. However, this view overlooks the possibility of an electronic contribution to the magnetic field, as proposed by Ref. \cite{transfer_to_electrons} and suggested by the large charge-to-mass ratio of the electron relative to the nucleus. The detailed mechanism by which nuclear angular momentum may transfer to the electronic structure, as well as a derivation of the resulting magnetic field strength, will be the subject of future work.
In the meantime, to contextualize the magnitude of phonon angular momentum observed in our simulations, it is instructive to compare it to a well-established excitation mechanism, by circularly polarized light (CPL). Prior experimental and theoretical studies have shown that CPL can induce substantial phonon angular momentum in certain materials, generating transient magnetic fields on the order of 1–100 T \cite{chir_phonons_optical,4f_para}. By constructing a simplified analytical model of CPL-driven excitation, we can benchmark the thermally induced phonon angular momentum observed in our simulations against these experimental values and assess whether thermal excitation could provide a competitive alternative.

An approximate benchmark model we use consists of an atom subject to an isotropic harmonic potential, with the Hamiltonian given by 
    \begin{equation}
        \begin{aligned}
            H_D&=\frac{1}{2}m\Dot{\Vec{x}}^2_j+ \frac{1}{2}m \omega_j^2 \Vec{x}^2_j.
        \end{aligned}
    \end{equation}
In contrast to Models A-C which involve chiral potentials subject to achiral driving forces, this potential $V_D=\frac{1}{2}m \omega_j^2 \Vec{x}^2_j$ is achiral and we take the driving force $\vec{F}_j(t)=q_j\vec{E}(t)$ to be a chiral, CPL pulse
\begin{equation}
\vec{E}(t)=(\hat{x}+i\hat{y})E_0e^{i\omega t} e^{-(t/\tau)^2},\label{driver}\end{equation}
where $E_0,\omega,$ and $\tau$ are the magnitude, frequency, and lifetime parameter respectively for the pulse. When subject to damping $\gamma_j$, motion of the atom $j$ is then governed by
\begin{equation}
\ddot{\vec{x}}_j+\gamma_j\dot{\vec{x}}_j+\omega_j^2\vec{x}_j=\frac{q_j}{m}\vec{E}(t).\label{differential}
\end{equation} 
Here $\gamma_i,\omega_i,$ and $q_i$ are the damping parameter and natural frequency, effective charge  respectively for atom $i$; $E_0,\omega,$ and $\tau$ are the magnitude, frequency, and lifetime parameter respectively for the pulse of light. For $t<\tau$ the complex valued solution, ignoring transients, is given by \begin{equation}
\Vec{x}_j(t)=\frac{(\hat{x}+i\hat{y})(q_j/m_j)E_0}{\sqrt{(\omega_j^2-\omega^2)^2+\gamma_j^2\omega^2}}e^{i(\omega t+\delta)},\end{equation}
which gives a maximal magnitude $u_j=q_jE_0/m\gamma_j\omega$ at resonance $\omega\rightarrow\omega_j$. Plugging this into the elementary formula for angular momentum $L_{z,j}=m_j u_j^2 \omega,$ we obtain \begin{equation}
L_{z,j}=\frac{q_i^2 E_0^2}{m_i \gamma_i^2\omega}.\label{model}\end{equation}

We can now compare the phonon angular momentum observed in our simulations to the phonon angular momentum in this CPL excitation model, Eq. (\ref{model}), using the values characteristic of the experiments in Ref. \cite{chir_phonons_optical} and \cite{4f_para}. Note that the value of the phonon angular momentum per atom in Model C, which contains the most realistic Hamiltonian of those examined in the main text, is approximately $L_{z,j}=\hbar/10$ (see Fig. 6(b)). We assume $q=0.1e$, $m_j=14$ g/mol, and $\gamma_j=0.1\omega$ and compare the two angular momentum magnitudes.

\begin{figure}[htbp]
        \includegraphics[width=240pt]{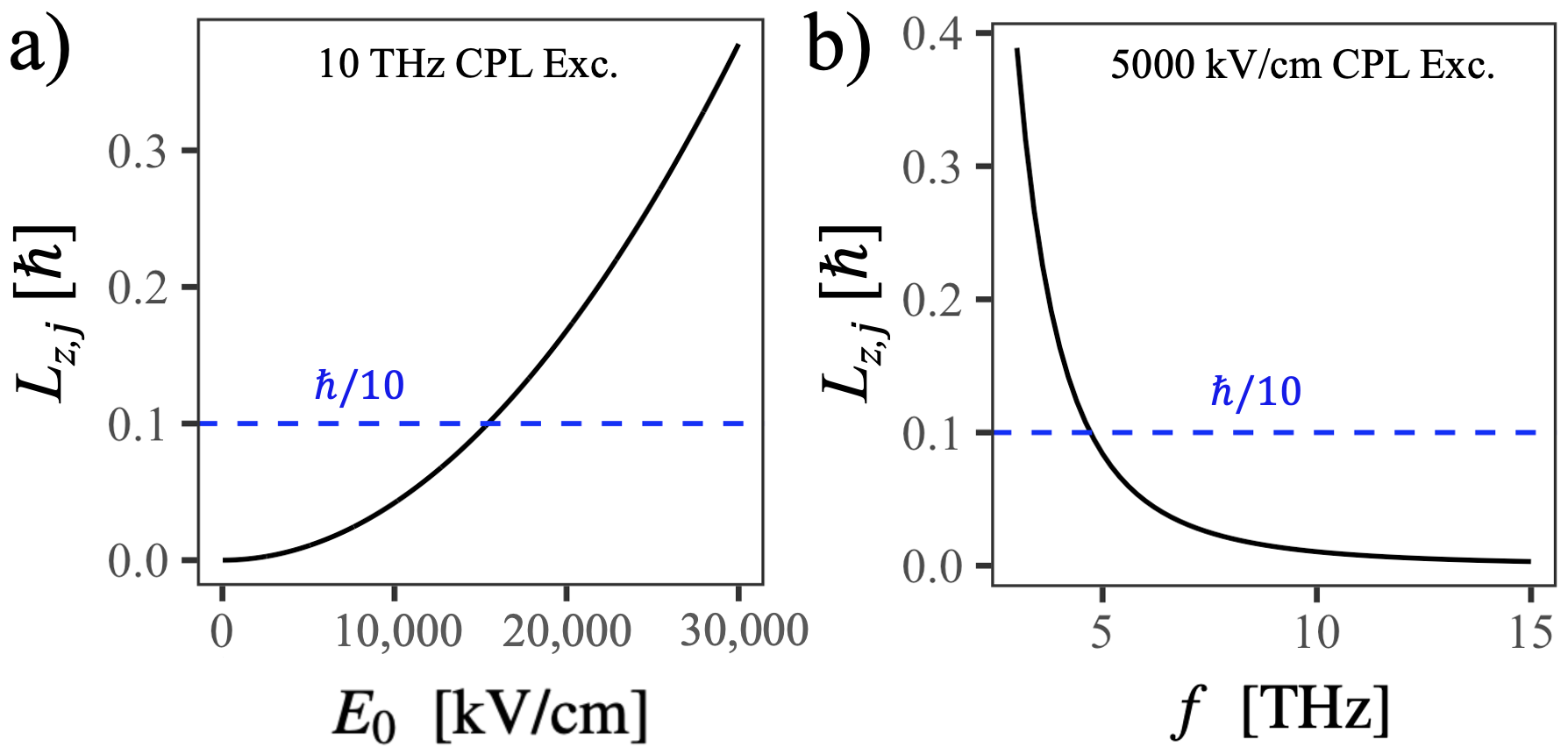}
        \caption{(a) The solid curve shows the excited phonon angular momentum predicted by Eq. (\ref{model}) for a 10 Thz pulse as a function of the pulse amplitude $E_0$. The dashed blue line shows the approximate value of the phonon angular momentum observed in our heat transport simulations and is included for comparison. (b) Same as (a), except the frequency $f=\omega/2\pi$, instead of the amplitude $E_0$, is varied. Here, the amplitude is held fixed at 5000 kV/cm. In both panels, we have set $q=0.1e,~m_i=14$ g/mol, and $\gamma_i=0.1\omega$.}
        \label{fig:AM_Z}
    \end{figure}

Figure 8 compares the prediction of Eq. \ref{model} to the $\hbar/10$ value observed in our simulations. In Fig. 8(a), we fix the CPL frequency and vary the amplitude while in Fig. 8(b), we fix the CPL amplitude and vary the frequency. Indeed, we see that the magnitude of the phonon angular momentum produced by the heat flux in our simulations is comparable to our estimates for the CPL excitation experiments at relevant parameter values \cite{chir_phonons_optical,4f_para}, and the magnitude can be further tuned by controlling the size of the applied thermal gradient and the frequency characteristics of the thermal driving.

\begin{table} 
\centering
\scriptsize
\vspace{\baselineskip}
\begin{tabular}{@{}>{\raggedright\arraybackslash}p{2.7cm}|
                >{\raggedright\arraybackslash}p{2.7cm}|
                >{\raggedright\arraybackslash}p{2.7cm}@{}}
\hline
\textbf{Feature} & \textbf{CPL Excitation} & \textbf{Thermal Excitation (This Work)} \\
\hline
Angular momentum per atom & \ensuremath{\sim \hbar/10} (extrapolated from Refs. \cite{chir_phonons_optical,4f_para}) & \ensuremath{\sim \hbar/10} (simulated in this work) \\
\hline
Duration & \ensuremath{< 1} ps; ultrafast transient & Steady-state \\
\hline
Coherence & Coherent (but decoheres quickly) & Incoherent \\
\hline
Experimental complexity & High: CPL source, tuning & Moderate: heat flux via contacts \\
\hline
Chirality origin & External---chiral perturbation & Internal---chiral material \\
\hline
Direction control & CPL handedness & Enantiomer or heat flux direction \\
\hline
Frequency resolution & Light frequency & Filtered thermal driving \\
\hline
System size scaling & Confined to optical penetration depth & Scales with system length $\propto L$ \\
\hline
\end{tabular}
\caption{\small Comparison between phonon angular momentum generation by circularly polarized light (CPL) and thermally induced angular momentum in chiral molecules, as predicted by our simulations.}
\label{tab:CPLvsThermal}
\end{table}

 A comparative summary of the two driving procedures is provided in Table~I. In the CPL excitation method, the material is achiral and the chiral agent is the external perturbation (i.e. the CPL), while in the thermal excitation method, the chiral agent must be the material itself and a the required perturbation is much less specialized. The CPL excitations are typically transient pulses ($<$1 ps), while the simulations presented in this work suggest that thermally excited phonon angular momentum can persist in the steady state. The directionality of the CPL-excited nuclear angular momentum is reversed by changing the handedness of the light's polarization, while the directionality of thermally-excited nuclear angular momentum can be reversed \textit{either} by changing the enantiomer of the substrate or the direction of the thermal gradient. A final difference we mention lies in the system-size scaling: While CPL excitations are confined to the optical penetration depth, the thermally excited counterpart scales roughly with the heat current, which remains approximately constant with system size. This implies that the total phonon angular momentum grows with system size, while the angular momentum per unit volume remains constant. Such scaling suggests that thermally driven nuclear angular momentum could be harnessed in macroscopic systems, as people have already observed the rigid body angular momentum generation in Einstein De Haas effect experiments.

Finally, it is interesting to note that we have modeled also a fourth system, Model D, consisting of a polypeptide consisting of 8 Alanine monomers (see Supplementary X for results and discussion of this Model D). In contrast to Models A-C described above, we observed only a subtle angular momentum response to the applied thermal gradient. We attribute this to the fact that the calculated thermal conductivity of this polymer was far lower than its polyethylene double-helical counterpart (129.28$\pm$4.43 vs. 2071.74$\pm$0.55 pw/K respectively; see Refs. \cite{ethan2023chain,Hadi} for calculation details), and that it showed much greater characteristics of diffusive (rather than ballistic) transport with its angular momentum dominated by the motion of the conformationally flexible side-chains. These observations are consistent with the view that the heat transport is more significant than the thermal gradient in determining the observed effect. 

\section{Conclusion}
We have studied theoretically and and computationally several chiral
models that exhibit angular momentum in the nuclear motion that underlies heat
transport. By examining both a harmonic helical chain models  and two molecular
models of a four-atom molecule on the one hand, and a polyethylene double helix on the other, we have demonstrated that phononic thermal conduction in such systems shows a significant angular motion character, highlighting a fundamental link between molecular symmetry
and heat-driven dynamics. Our principal findings are: 
\begin{quote}(i) The the total angular momentum
generated in such chiral structures while carrying thermal currents is an average over frequency dependent responses that can have different magnitudes and, arguably more important, sign in different parts of the spectrum.\\
(ii) The directionality of the angular momentum is inherently coupled with the direction of heat flux, suggesting a role of structural chirality in guiding energy
transfer.\\
(iii) The generated angular momentum is nearly linear in the heat current. In the linear response regime for a system of fixed length, it is linear in the overall temperature fall of the system, but not necessarily in the local temperature gradient. \\(iv) Anharmonic effects can modify the linear dependence of the angular momentum on the
thermal imbalance.\\
(v) While a quantitative dependence of the magnitude of the genetrated
angular momentum on the system chirality could not be determined in realistic models, such
dependence was clearly observed in our simulations as evidence in correlation observed
between the angular momentum and the imposed dihedral angle in the 4-aton model.\\
(vi) The magnitude of the generated angular momentum is comparable to that which is generated by exciting chiral phonons with
circularly polarized light.
\end{quote}

Our results highlights the robust nature of angular momentum generation across
different scales and system configurations, whether it is in a small, four-atom chain or a larger, more complex helical structure. Recent observations that implicate phononic angular
momentum with magnetic phenomena point to the possibility of realizing heat-transfer induced magnetic response or magnetic control of heat transfer phenomena.

Further work is needed for considering the the contributions of electrons to such phenomena. Our current calculations are limited to
the dynamics in the nuclear subsystem and considering the electron-phonon interactions and the possibility that nuclear angular momentum can be transferred to electronic orbital and spin dynamics may lead to better assessment of the interplay between thermal transport and magnetic phenomena and possible implications for the CISS effect.

Finally, as noted above, the correlation between the directions of the heat current and the generated angular momentum suggest the possibility that molecular chirality may provide a route for controlling the directionality of vibrational energy transfer. This issue, already
suggested in Ref. \cite{chen2022diodetheory}, will be addressed in a forthcoming article.
\\
\section*{ACKNOWLEDGMENTS}
The research of A.N. is supported by the Air Force Office of Scientific Research under award number FA9550-23-1-0368. E.A. acknowledges the support of the University of Pennsylvania College Alumni Society Undergraduate Research Grants. We thank Claudia Climent for many helpful discussions, and Mohammadhasan Dinpajooh for valuable technical assistance.

\section*{AUTHOR DECLARATION}
\subsection*{Conflict of Interest}
The authors have no conflicts to disclose.

\section*{DATA AVAILABILITY}

The data that support the findings of this study are available from corresponding authors upon reasonable request. Sample input files for Model C are available at \url{https://github.com/eabes23/polymer_twist/}.

\bibliography{reference}
\end{document}